# Nudged elastic band calculations of stacking and dislocation pathways in diamond

Chunxia Chi[1], Hairui Ding[1], Xiang-Feng Zhou[2], Xiao Dong[1,*]

[1]*Key Laboratory of Weak-Light Nonlinear Photonics and School of Physics, Nankai University, Tianjin 300071, China*

[2]*Center for High Pressure Science, State Key Laboratory of Metastable Materials Science and Technology, School of Science, Yanshan University, Qinhuangdao 066004, China*

Diamond, the hardest natural crystal, has attracted significant attention for its plasticity, which is reported to be determined by its stacking faults. Studies mainly focused on one-dimensional linear pathways in stacking transitions, neglecting its transverse freedom on the main slip plane. However, in an actual stacking procedure, stacking faults can follow curve line along the slip plane rather than constrained to straight lines. In this study, using *ab initio* calculations, we mapped the $\gamma$-surface, defined as the landscape of generalized stacking fault energies, along the weakest direction of the {111} orientation in diamond. We then applied the Nudged Elastic Band (NEB) method to determine the minimum energy paths, finding significantly reduced stacking energy barriers compared to previous reports (for the glide-set, our energy barrier is only one-third of that for the traditional direct path). Our calculations reveal that the glide-set can round its high-energy peak, with a lower energy barrier within the entire stacking plane than the shuffle-set. By employing the NEB method, we have constructed the minimum energy path (MEP) for both the stacking and dislocation procedures. Our results provide new insights into the plasticity and stacking faults of diamond, advancing the understanding of superhard carbon material transition, especially the diamond under shear stress.

# I. INTRODUCTION.

As a typical covalent crystal and the hardest known material, diamond has garnered extensive attention, particularly due to its brittleness [1-8]. Unlike ductile metals [9-11], the high bond energies and strong directionality of covalent bonds in diamond enhance its strength but reduce its plasticity. When diamond fractures—especially under shear stress—it tends to break into pieces due to broken bonds, which are difficult to repair. Theoretically, dislocations are inextricably related to stacking faults. The multiplication and decomposition of dislocations correspond to local stacking faults. Stacking fault energy is often used to characterize the slip resistance of dislocations and plays a crucial role in this fracture process, not only in diamond but also in other covalent superhard materials, attracting significant scientific interest.

Plastic deformation is inherently a multiscale process influenced by factors such as temperature [12, 13], hydrostatic pressure [14], and stress, particularly shear stress [15, 16]. Generalized stacking fault energy (GSFE) is often used as a standard measure of crystal plasticity under various shear deformation modes, based on the Peierls-Nabarro model [17, 18]. The GSFEs are calculated using the equation $(E_f - E_p)/A$, where $E_f$ represents the energy during stacking faults as a function of stacking atomic coordination, $E_p$ is the energy of a perfect crystal (i.e., the ground state), and $A$ is the area of the slipping surface.

Recently, significant research has focused on plastic deformations and stacking faults in materials with low stacking fault energies, such as metals and other high-ductility materials [19-24]. These studies have shown that GSFEs, particularly the energy barriers, are a decisive factor in determining ductility and brittleness [25]. Similarly, for strong covalent crystals like diamond, reducing the stacking fault energy barrier can significantly enhance ductility, addressing the challenge of balancing high strength with plasticity.

While stacking fault energy barriers are critical, the modeling and calculation methods for stacking fault pathways and energy barriers require further refinement. The key problem is that stacking occurs in a free two-dimensional plane, not along a one-dimensional straight line. The γ-surface, defined as the landscape of GSFEs along the whole stacking fault freedom degree, is important for determining pathways and barriers. In metals, the distinction is minor due to their low barriers, but in covalent superhard materials, the differences can be substantial due to their high bond energies. In 1999, Schoeck et al. analyzed the γ-surface of $Ni_3Al$, identifying two-dimensional stacking fault pathways connecting saddle points with local minima by studying GSFEs across this plane [24]. Subsequent research has corroborated stacking fault dissociation within slip planes in metals [19-24, 26-28], and similar phenomena have been observed in covalent crystals such as Si [29] and BN [30]. In 2020 [3], Nie et al. directly observed the plasticity of diamond at room temperature experimentally, using the method that involved driving diamond nanoneedles under compressive loading.

Therefore, the stacking fault path of diamond along the entire γ-surface, and its dissociation phenomenon, warrant further investigation rather than limiting the study to a single crystallographic direction. Previous research, calculated direct paths[13, 31] or considered direct paths on a whole GSFE landscape[32], has focused solely on one-dimensional pathways, neglecting the possibility of slippage along curved paths within the planes, which more accurately reflects reality. According to the close-packed γ-surface of diamond, which is the main slip plane in face-centered cubic (fcc) ceramics and similar structures [33-39], the sliding potential energy landscape of the γ-surface forms an uneven terrain of peaks and valleys. Thermodynamically, diamond's sliding on the γ-surface must follow the minimum energy path (MEP) [40], not a straight crystal direction. According to Henry Eyring's transition state theory (TST) [41], the energy barrier along the MEP corresponds to the saddle point on the potential energy landscape of the γ-surface, and the structure at the saddle point represents the transition state of diamond from one state to another. These intermediate states, such as saddle points and local minima, as essential aspects reflecting the stacking mechanism, are pivotal points in dissecting and analyzing the stacking pathway.

To identify the MEP on the γ-surface, among the numerous theories and methods explored in previous studies [29, 42-54], we employed the Nudged Elastic Band (NEB) method [55, 56] for its low computational cost, algorithmic stability, strong directivity, and successful applications in searching for transitional phases in covalent crystals such as borides [57] and carbides [58]. This method is employed to find the MEP between determinant initial and final states on a given potential energy surface. Introduced by Henkelman et al. [59, 60], the NEB method optimizes the path on the potential energy landscape by combining image points and springs. It serves as an effective tool for optimizing the pathways, connected the fixed initial and final states by several virtual elastic springs and image points on the high-dimensional energy landscape. This arrangement not only prevents all images falling to local minima but also ensures their uniform distribution between the initial and final states, thereby iteratively converging towards the MEP.

In this study, we used first-principle calculations and density functional theory to calculate the GSFE of diamond, independent of temperature, on two inconsistent sets of slip planes for both the glide-set and shuffle-set, corresponding to close and wide spaces across the entire γ-surface, respectively. Based on these GSFE landscape, we utilized the NEB method to obtain the MEPs connecting stable states through the $\langle 1\bar{1}0 \rangle$, $\langle 11\bar{2} \rangle$, and $\langle 20\bar{2} \rangle$ directions on the γ-surface of diamond. Optimized by the NEB method, the direct path along a specific crystallographic orientation converged and decomposed into several consecutive segments, each passing through a local minimum on the GSFE landscape. Each segment traverses its respective saddle point, representing the stacking fault dissociation process. Our work confirms that stacking fault dissociation occurs even in diamond, the hardest natural crystal.

## II. METHODS.

The first-principles calculations were performed using density functional theory, with electron exchange-correlation interactions treated via the all-electron projector augmented wave (PAW) method [61, 62] and parametrized by local density approximation (LDA) functional [63, 64], as implemented in the Vienna *ab initio* simulation package (VASP) [65]. A conjugate-gradient algorithm [66] was used to relax the structures containing stacking fault dislocations on γ-surface during the calculation of GSFEs, allowing atomic relaxation only in the direction perpendicular to the slip planes. Throughout the calculations, we used a plane-wave kinetic energy cutoff of 800 eV and a *k*-point mesh of 0.2 Å$^{-1}$ generated at a center of a Γ point to sample the Brillouin zone, with a high convergence criterion of the total energy and all the forces on atoms set to 10$^{-8}$ eV and 0.001 eV/Å respectively. Specially, we used the density of 100 *k*-points taken in each *k*-path in the self-consistent and energy band computation. *k*-paths start from the point Γ(0 0 0) and path sequentially through B(0.5 0 0), W(0.5 0.5 0), F(0 0.5 0) and finally back to Γ(0 0 0).

The diamond structure consisted of 48 atoms and a 15 Å vacuum layer, with lattice vectors oriented along ⟨11$\bar{2}$⟩, ⟨1$\bar{1}$0⟩, and ⟨111⟩ directions. The atoms were arranged in 24 layers along the ⟨111⟩ direction, where stacking fault dislocations on the γ-surface were generated by slipping the top 12 (shuffle-set) or 13 (glide-set) layers over the lower layers along lattice vectors formed by various combinations of ⟨11$\bar{2}$⟩ and ⟨1$\bar{1}$0⟩. A slip mesh of 31 × 31 (961 points) was employed to uniformly divide the γ-surface of diamond and generate stacking fault dislocations, enabling the calculation of the GSFE landscape.

We developed a set of NEB iterative programs based on Python. The step size for each image movement was set to one-thousandth of the lattice constant ($a_0$), with a convergence criterion of 0.001 eV/Å for the resultant force and a spring constant of 10 eV/Å$^2$ for the elastic band. For initial image paths, we used linear interpolation, sampling 15 to 21 images along the direct path at equal intervals. After achieving convergence in the first iteration, we performed successive linear interpolations between existing images using intervals of 0.05 $a_0$, 0.04 $a_0$, 0.03 $a_0$, 0.02 $a_0$, and 0.01 $a_0$, iterating and converging at each stage. This process is repeated until the number of images before and after the linear interpolation becomes the same. The corresponding numbers of images for shuffle-set ⟨11$\bar{2}$⟩, ⟨1$\bar{1}$0⟩, and ⟨20$\bar{2}$⟩ at the initial sampling and final convergence stages are (15→68), (15→46), and (21→71) respectively. Similarly, for glide-set ⟨11$\bar{2}$⟩, ⟨1$\bar{1}$0⟩, and ⟨20$\bar{2}$⟩, the number of images at the initial sampling and final convergence stages are (19→74), (15→47), and (21→79) respectively.

### III. RESULTS & DISCUSSION

As shown in Fig. 1, the γ-surfaces for the shuffle-set and glide-set along the {111} direction in diamond reveal notable differences. Our γ-surfaces is similar to previous work [32]. The glide-set landscape is steeper than that of the shuffle-set, with the latter displaying lower peaks. However, the saddle points in the glide-set have lower energies than those in the shuffle-set, indicating that the glide-set landscape is rougher and allows only narrow, winding transition paths during the stacking process. To analyze the configuration and differences of the γ-surfaces of diamond, Fourier transforms were conducted on the GSFEs of the glide-set (Fig. 1(e)), shuffle-set (Fig. 1(d)), and aluminum (Fig. 1(f)) for comparison. The GSFEs configuration of diamond's shuffle-set resembles that of metallic elements like aluminum, whereas the glide-set displays a distinct energy landscape due to the presence of additional local minima structures. Subsequently, the intensities of the resultant patterns were transformed using natural logarithms and depicted in Fig. 1. The most significant difference between the shuffle and glide-sets is that the glide-set exhibits enhanced signals at second-order patterns, corresponding to wave vectors, such as ⟨2$\bar{1}\bar{1}$⟩, which result from wave modes induced by these local minima. Structural analysis suggests these local minima stem from a hybrid diamond structure, combining hexagonal and cubic arrangements, which will be explored in detail in the pathway discussion. These unique local minima structure within the glide-set introduces additional pattern signals in reciprocal space, enriching the energy composition of the reciprocal space with novel modes.

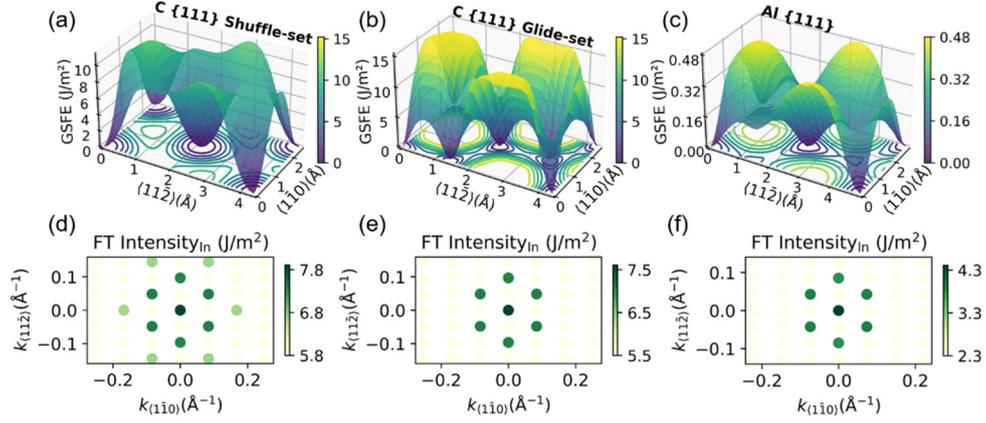

**Fig. 1**. GSFEs and Fourier Transform (FT) intensity for diamond's γ-surface at shuffle-set (a, d) and glide-set (b, e), and Aluminum at {111} plane (c, f).

Using the NEB method, we determined the true stacking pathways and energy barriers along the $\langle 1\bar{1}0 \rangle$, $\langle 11\bar{2} \rangle$, and $\langle 20\bar{2} \rangle$ directions. As shown in Fig. 2, these paths break into several segmented, curved trajectories defined by a few local minima. For glide-sets, due to the triple rotational symmetry, the pathway from a perfect crystal to a hybrid diamond forms a hexagonal network connecting each perfect crystal point in the γ-surface. Thus, each stacking pathway from one perfect crystal point to another in the glide-set can be decomposed into several basic segments connecting the perfect crystal and hybrid diamond states. Conversely, the basic segments for the shuffle-set run from one perfect crystal point to another, along the red lines shown in Fig. 2(d-f). This hexagonal network is critical for the following MEP discussion.

The MEPs of glide-set corresponding to these various slip directions are outlined below.

$$\langle 1\bar{1}0 \rangle = \frac{1}{3}\langle 1\bar{2}1 \rangle + \frac{1}{3}\langle 2\bar{1}\bar{1} \rangle \tag{1}$$

$$\langle 11\bar{2} \rangle = \frac{1}{3}\langle 11\bar{2} \rangle + \frac{1}{3}\langle 2\bar{1}\bar{1} \rangle + \frac{1}{3}\langle 11\bar{2} \rangle + \frac{1}{3}\langle \bar{1}2\bar{1} \rangle \tag{2}$$

$$\langle 20\bar{2} \rangle = \frac{1}{3}\langle 11\bar{2} \rangle + \frac{1}{3}\langle 2\bar{1}\bar{1} \rangle + \frac{1}{3}\langle 11\bar{2} \rangle + \frac{1}{3}\langle 2\bar{1}\bar{1} \rangle \tag{3}$$

Compared to the direct path, these MEPs actively bypass the peak points on the GSFE landscape by diverting toward saddle points and breaking into multiple trajectories connected through local minima, reducing energy barriers. Notably, the GSFE curve along the MEP exhibits a significantly lower energy barrier, particularly for the glide-set slip layer. This makes the glide-set more energetically favorable than the shuffle-set. The reduced energy barrier simplifies the stacking process, providing a more realistic and likely pathway on this landscape. Because the MEPs circumvent high energy barriers, the energy barrier along the entire stacking pathway becomes equivalent to the first barrier encountered. This underscores that once dislocation is initiated on the γ-surface of diamond, any subsequent activation is solely contingent upon the height of this initial barrier. No additional energy, beyond that required to overcome this initial energy barrier, is necessary. Higher energy barriers encountered later in the process simply influence the dislocation's slip path without augmenting the slip's difficulty.

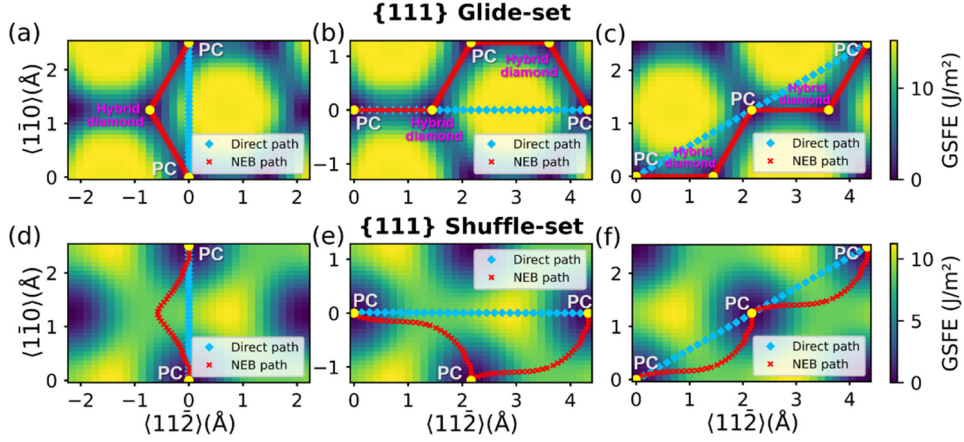

**Fig. 2**. Slipping paths for direct paths (blue solid rhombus) and NEB paths (read crosses) in $\langle 1\bar{1}0 \rangle$ (a, d), $\langle 11\bar{2} \rangle$ (b, e), and $\langle 20\bar{2} \rangle$ (c, f) at glide-set (top) and shuffle-set (bottom) on γ-surface of diamond. The yellow solid balls represent positions with formations for the perfect crystal (PC) and hybrid diamond.

Numerous local minima and saddle point states were observed within the stacking fault dislocations on the γ-surface of diamond. For instance, when stacking faults are located halfway along the lattice constants of the $\langle 1\bar{1}0 \rangle$ and $\langle 11\bar{2} \rangle$ directions, this positioning is equivalent to having a stacking fault spanning an entire lattice period along the $\langle 10\bar{1} \rangle$ direction, resulting in a structure identical to a perfect crystal (PC), which coincides with valleys in the GSFE landscape.

As depicted in Fig. 2, the stacking faults within the γ-surface of diamond are constructed from several basic routes from perfect crystal to hybrid diamond, governed by the symmetry of the diamond structure. These routes connect to a hexagonal network to reach every thermal stable point, i.e. perfect crystal or hybrid diamond. For the glide-set, the $\langle 1\bar{1}0 \rangle$ direction requires at least two routes, while the $\langle 11\bar{2} \rangle$ and $\langle 20\bar{2} \rangle$ directions require at least 4 routes from one pure crystal state to another pure crystal state. Most previous studies [30, 31, 67] do not consider the pathway along $\langle 11\bar{2} \rangle$ after reaching the hybrid diamond state, as it is believed that the energy barrier was too high to overcome in an actual phase transition. However, our newly discovered pathway offers an alternative by bypassing the high energy barrier rather directly crossing it. This approach reduces the energy barrier to 5.87 J/m²—about one-third of the direct path barrier —making it possible to complete the remaining stacking transition to a perfect crystal. Therefore, for any complete stacking pathway, stacking from one pure crystal to another in the glide-set is possible by following the NEB pathway with dislocation decomposition. Upon comparing the pathway of $\langle 1\bar{1}0 \rangle$, $\langle 11\bar{2} \rangle$ and $\langle 20\bar{2} \rangle$, the $\langle 1\bar{1}0 \rangle$ pathway precisely corresponds to a basic slip unit. This pathway is not only relatively shorter than the other pathways, but also is an initial necessary step for other pathways, such as the first $\langle 1\bar{1}0 \rangle$ and $\langle 10\bar{1} \rangle$ stacking in the beginning of slip movements on $\langle 20\bar{2} \rangle$ and $\langle 11\bar{2} \rangle$ direction. Consequently, the slip along the $\langle 1\bar{1}0 \rangle$ pathway and its equivalent $\langle 10\bar{1} \rangle$ direction occurs more readily and is more prevalent compared to the slip along the $\langle 11\bar{2} \rangle$ and other orientations. This finding is consistent with prior experiment and analysis of diamond dislocation [3].

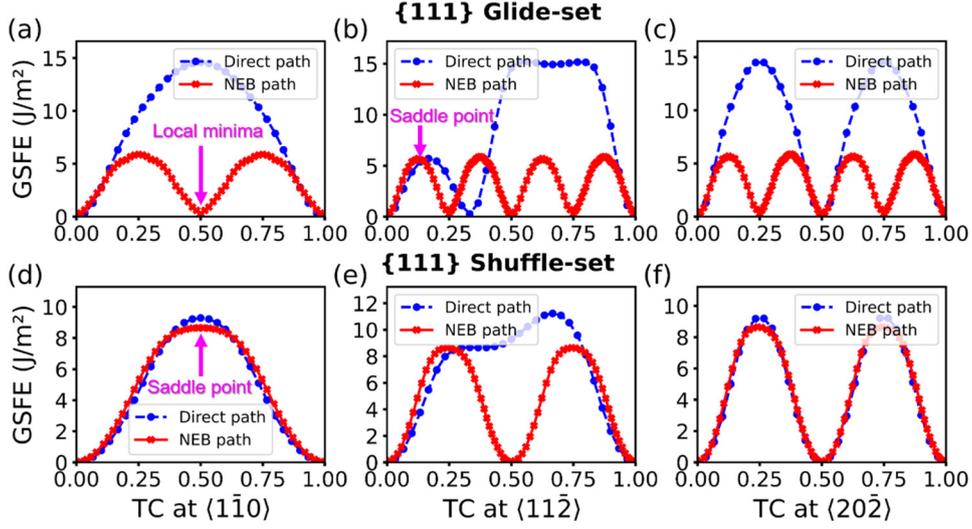

**Fig. 3**. GSFEs-Transition coordinate (TC) of direct paths (blue dotted lines) and NEB paths (red crossed lines) in ⟨1$\bar{1}$0⟩ (a, d), ⟨11$\bar{2}$⟩ (b, e), and ⟨20$\bar{2}$⟩ (c, f) at glide-set (top) and shuffle-set (bottom) on γ-surface of diamond.

For the stacking along the ⟨1$\bar{1}$0⟩ direction in the shuffle-set, a small shift of 0.57 Å along the ⟨11$\bar{2}$⟩ direction occurs at the barrier point compared to the direct path, breaking the symmetry between the upper and lower layers. This symmetry breaking slightly decreases the energy barrier by 0.64 J/m$^2$, bringing it to 8.65 J/m$^2$. Similar to the glide-set, this pathway from pure crystal to pure crystal forms a triangular network to reach every pure crystal point on the γ-surface. For example, the stacking along the ⟨11$\bar{2}$⟩ direction can decompose into two routes, ⟨01$\bar{1}$⟩ and ⟨10$\bar{1}$⟩, significantly reducing the energy barrier from 11.24 J/m$^2$ for the direct path to 8.65 J/m$^2$ for the NEB path, as shown in Fig. 3(e).

The basic route of the glide-set has a lower energy barrier than that of the shuffle-set. Moreover, due to the symmetry and construction of the γ-surface, every glide-set pathway from one pure diamond point to another has a lower energy barrier than the shuffle-set. These findings, based on the NEB method, differ significantly from previous studies [31, 67] which suggested that the shuffle-set has a lower energy than glide-set along the direct path. However, several theoretical simulations of diamond under dislocation [7] and experimental studies on silicon (which shares the diamond crystal structure) under shear [68] show that the glide-set is more favorable in dislocation and stacking procedures. Our research using NEB paths provides new insights into the stacking pathways for glide-set, shuffle-set and their relative energy barriers. The structures and properties of key images, including saddle points and local minima are shown in Fig. 4(a) and Fig. 5(a-c). The stacking fault transitions encompass fracture (saddle point in the shuffle-set), nanoscale phase transitions (local minima in the glide-set), and semi-metallization (saddle point in the glide-set), which are further discussed below.

1. Fracture in the shuffle-set transition.

As shown in Fig. 4(a), at the energy barrier point for the shuffle-set transition, the interface between the upper sliding layers and the lower substrate is torn, with the carbon-carbon bond distance increasing to 2.41 Å. In Fig. 4(b), the band-decomposed charge density near the Fermi level reveals a weakening of the bonding and electronic overlap between the upper and lower layers, with the wave function breaking into two single dangling *p*-wave bonds. This suggests that the fracture occurs not only at the atomic structure and bond length level but also at the electronic distribution level, which explains why the shuffle-set has a higher energy barrier than the glide-set transition.

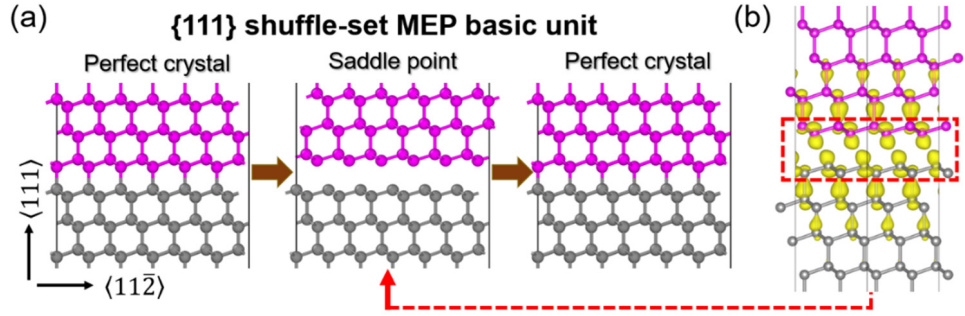

Fig. 4. (a) MEP basic unit for shuffler-set. (b) Isosurfaces of charge density (yellow parts) for a structure at saddle point through {111} shuffle-set MEP. The sliding and the substrate are colored as pink balls and grey balls respectively.

2. Hybrid diamond of local minima in glide-set transition.

It is counterintuitive that shuffle-set stacking faults with wider interlayer spacings have a higher energy barrier than the glide-set. As discussed above, glide-set indeed has high peaks on the $\gamma$-surface. However, in the stacking fault process, nature supplies several valleys (hybrid structure with both cubic and hexagonal diamond) on the $\gamma$-surface to decrease the energy barriers, which is the main difference between shuffle-set and glide-set. As shown in Fig. 5(a-c), this hybrid diamond structure acts as a local minimum, connecting the initial and final states along the MEP for the glide-set. Our calculations show the grain boundary contains two layers of hexagonal diamond (Lonsdaleite), with a thickness of 4.65 Å. A similar coherent interface is recently reported to exist in the shear procedure of diamond [69]. Since this twin structure represents a local minimum, it can persist under normal conditions, explaining the formation of nanoscale hexagonal diamond and the migration of nanotwinned diamond boundaries under shear stress [70, 71].

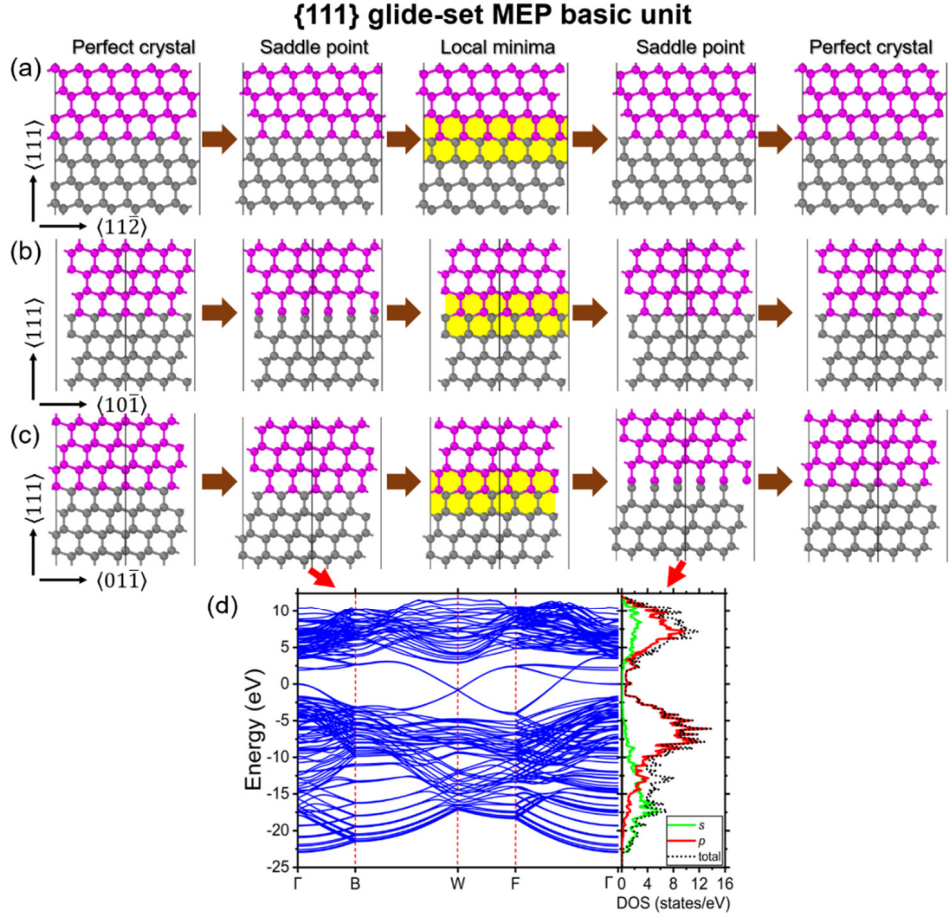

Fig. 5. (a-c) MEP basic unit for glide-set. The slip system and the substrate are colored within (a-c) as pink balls and grey balls respectively. The yellow parts show the hexagonal diamond (Lonsdaleite) of the twin boundary in the glide-set MEP. (d) Energy bands and density of states (DOS) for a structure at saddle point through {111} glide-set MEP.

3. Semi-metallization at the saddle point of glide-set transition.

In contrast to the fracture in the shuffle-set transition, the glide-set maintains carbon-carbon bonding between the upper and lower layers. With a sliding along the MEP for glide-set, the original four-fold $sp^3$ carbon atoms at the grain boundary transition to the three-fold $sp^2$ hybridization, with only one bond breaking. The saddle point structure is a typical gradia [58, 72] with both $sp^3$ (diamond part) and $sp^2$ (graphite part) hybridization. This change in bonding state and hybridization mode causes diamond to transition from an insulator to a semi-metal with a Dirac nodal line, similar to other carbon allotropes [73-81]. We calculated the energy bands and density of states (DOS) of this structure in Fig. 5(d) and found that the bandgap closes, resulting in a semi-metallic interface. At the Fermi level, the electrons in the $p$-orbitals dominate. This saddle point appears twice in the sliding path of the glide-set along the $\langle 1\bar{1}0 \rangle$ direction, with different crystal orientations as shown in Fig. 5(b-c) due to the three-fold rotational symmetry of the diamond {111} axis.

For a detailed visualization of the stacking fault sliding process along the MEP, please refer to videos

1-6 in the supplementary materials.

The NEB method offers greater scalability in the algorithm, as it requires only the atomic forces of images, which are relatively easy to obtain from first-principles calculations. This enables us to determine the global energy barrier and pathways from several NEB steps, requiring only the initial state and final stacking states. Using the NEB method eliminates the need for expensive calculations of the γ-surface, which often includes excessive, non-essential information such as peak energies and detailed energy landscapes.

## IV. CONCLUSIONS

In summary, using first-principles calculations, we have determined the γ-surface and stacking pathways for the shuffle-set and glide-set along the {111} orientation in diamond. Unlike previous studies based on direct, straight stacking paths, our pathways on the γ-surface are two-dimensional curves that account for the stacking dissociation phenomenon. The energy barriers of these new pathways are much lower than those of the direct path, and the MEP for the glide-set shows a lower energy barrier than the shuffle-set, indicating a preference for glide-set sliding. Furthermore, an analysis of key images at the saddle points and local minima in the transition pathways reveals several interesting phenomena, including fracture (saddle point in the shuffle-set), the formation of nanoscale hexagonal diamond (local minima in the glide-set), and semi-metallization (saddle point in the glide-set). These results provide new insights into the plasticity and stacking faults of diamond, contributing to a better understanding of the formation and transitions of hybrid diamond and graphite-diamond carbon materials, especially under shear stress.

## ACKNOWLEDGMENTS

This work was supported by the NSFC (Grants No. 92263101 and 12174200). The calculations were performed and supported by Tianhe II in Guangzhou.

___________________________

*xiao.dong@nankai.edu.cn